\documentstyle[12pt,epsf,oldlfont]{article}
\begin{document}

\def\singlespacing{\baselineskip=12pt}
\def\doublespacing{\baselineskip=24pt}
\pagestyle{empty}

\noindent
July 21, 1997  \hfill {\bf MC-TH-97/12}

\vspace{2cm}

\begin{center} 
\begin{large}

{\bf  FROM DEEP INELASTIC SCATTERING TO PHOTOPRODUCTION: A UNIFIED APPROACH}

\bigskip

{\em G. Kerley and G. Shaw}

\end{large}

{Department of Physics and Astronomy, Schuster Laboratory\\University of 
Manchester, Manchester M13 9PL, U.K.}
\end{center}

\vspace{1cm}

\begin{center}
{\bf ABSTRACT}
\end{center}

\noindent
The  strikingly different high energy behaviours of real photoabsorption
cross-sections with $Q^2 = 0$ and the low $x$ proton structure function  at 
large $Q^2$ are studied from a laboratory
frame viewpoint, in which the $x$ and $Q^2$ dependence reflects the
space-time structure of the interaction. This is done
 using a simple model which incorporates hadron dominance, but attributes the 
striking enhancement observed at HERA at very low $x$ and high $Q^2$ to
contributions from heavy long-lived fluctuations of the incoming photon.
Earlier published predictions of the model for the then unknown behaviour of 
the
structure  function at small $x$ and intermediate $Q^2$ are shown to be
strikingly confirmed by recent experimental data. A simultaneous analysis
of real photoabsorption data and  structure function data for 
 $0 \le x < 0.1$ and $ 0 \le Q^2 \le 15 \; GeV^2$ is then reported. 
An excellent fit is obtained, with all  parameters in the restricted ranges 
allowed by other physical requirements.

\newpage
\pagestyle{plain}
\pagenumbering{arabic}

\section{ Introduction}

At high energies, the real photoabsorption cross-section
$$
\sigma_{\gamma p}(\nu, Q^2 = 0) \approx a_P \nu^{\alpha_P - 1}
+ a_R \nu^{\alpha_R-1}
$$
is characterised by the intercept $\alpha_P \approx 1.08$ of the
\lq \lq soft'' pomeron familiar from hadron scattering
\cite{DL1}, together with an effective contribution from lower
lying Regge trajectories with $\alpha_R \approx 0.55$. The same behaviour
characterises the proton structure function in the intermediate $x$-region
$ 0.02 \le x \le 0.1$, but  experiments at
HERA \cite{H193,ZEUS93} in 1993 first observed the  much sharper 
 rise in the 
structure function at small $x < 0.02$ and large $Q^2 > 8 \; GeV^2$   
associated with the 
\lq\lq hard'' pomeron. Here we focus on the transition
between these contrasting behaviours in the intermediate $Q^2$ region
 $ 0 <  Q^2 < 8 \; GeV^2$ using an approach suggested by one of 
us \cite{Shaw93}. This was used to predict the  behaviour 
at intermediate $Q^2$ in 1994 \cite{MS95a}, but   there was 
no data with which to confront the predictions. Subsequently the
experimental situation has been transformed by precise data both at high $Q^2$ 
and in the hitherto unexplored region of low $x$ and intermediate $Q^2$.
\cite{H196,ZEUS96,E665}. Here we examine whether the new data at intermediate
 $Q^2$ agree with the prior predictions of the model; and whether its greater
precision and its extension to smaller $x$-values enable the  intercept of 
the  hard pomeron to be determined more precisely
by the data.

The model used is described in the next section. However it will  be useful
to first summarise the underlying approach \cite{Shaw93}.
This   adopts the laboratory
frame viewpoint  typical of hadron dominance  models, and emphasizes the 
role of the
coherence length
\begin{equation}
l = \frac{2\nu}{m^{2}\,+\,Q^{2}} = \frac{1}{M x}  
\frac{1}{1 + m^2/Q^2} >  \frac{1}{M x} \; ,
\label{i2}
\end{equation}
which represents the typical distance travelled by a vacuum fluctuation 
of the photon of mass $m$. Our assumptions on the nature of the dominant states
at different coherent lengths are summarised in Figure 1. At short coherence 
lengths, $ l < l_0 \approx 1$ f,
corresponding to large $x$ values $x  > 0.1$,  we assume that  
these states are essentially bare $\bar{q}q$ pairs and a hadronic component
of the photon has not developed. At moderate coherence lengths,             
they are assumed to have the single hadron-like behaviour expected for 
constituent $\bar{q}q$ pairs or
vector mesons. However at very long coherence lengths, the photon fluctuations 
can eventually develop into the hadronic final states observed 
in \( e^{+}e^{-} \) annihilations\footnote{For an attempt to understand the 
role of the coherence length from a more fundamental dynamical viewpoint,
see Del Duca, Brodsky and
Hoyer \cite{Brodsky92}.}. For low masses, these states can be 
approximated by a sum of vector 
mesons; but for high masses complicated  jet-like final states 
are observed.
The idea is that the new 
phenomena at very small $x$ are associated with complicated jet-like states
 which only play a
role for masses $m$  and coherence lengths $l$ which are greater than 
some critical values $ m > m_J$ and
$ l > l_C$. By (\ref{i2}) this  implies that they are confined to energies
\begin{equation}
\nu > \nu_C \equiv m_J^2 l_C / 2 \hspace{2cm} ( Q^2 = 0)
\label{i3}
\end{equation}
in photoproduction, and to the kinematic region
\begin{equation}
x < x_C(Q^2) \equiv \frac{1}{M l_C} \frac{Q^2}{Q^2 + m_J^2}
\label{i4}
\end{equation}
for structure functions. The remaining states with $ m < m_J$ and/or
$ l < l_C$  will completely
dominate outside this region. They are included using conventional hadron 
dominance ideas  which emphasize the link
between real and virtual photons and provide a natural framework for describing
the onset of scaling in the region\footnote{In hadron 
dominance models which give
approximate scaling, the bulk of the contributions for a given $Q^2$ comes from
intermediate states with $0 < m^2 < n \, Q^2$, with $n$ a small integer.  
The average value is of order $m^2 \approx Q^2$, for which 
the coherence length $ l \approx 1/2Mx$ by (\ref{i2}). }  $ x_C <x < 0.1$ .
For example, they give a good description of the 
transition between real photoabsorption and deep inelastic scattering
data on protons in the intermediate $x$-region $ 0.02 \le x \le 0.1$ 
\cite{Shaw93, MS95b}; and they
successfully predict the observed shadowing behaviour for real photoabsorption
and deep inelastic scattering on nuclei in the same $x$ 
region \cite{Shaw93a, Shaw89, Arneodo94}.

\section{A simple model}                     
                                    
In the hadron dominance model\footnote{For  reviews, see for example 
\cite{Shaw93} and
\cite{Review} and references therein.}, the total 
cross-section for 
photoabsorption is given by an expression of the form
\begin{equation}
\sigma_{\gamma p}(\nu,Q^2) = \int dm^2 \int dm^{\prime 2}
\frac{\rho (m,m^\prime,s)}{(m^2 + Q^2)(m^{\prime 2} +Q^2)} 
\label{fp1}
\end{equation}
corresponding to Figure 2. Conventional hadron physics and duality with the
parton model both require large  \lq \lq diagonal'' $m = m^{\prime}$ and 
large \lq\lq off-diagonal'' $m \ne m^{\prime}$ components to be present; 
and the
latter must be explicitly incorporated if the approximate scaling behaviour
of the shadowing for deep inelastic scattering on heavy 
nucei is to be understood \cite{Shaw93a, Shaw89, Arneodo94}. However, these 
off-diagonal contributions are dominated by terms in which 
the two mass values $m$ and 
$m^{\prime}$ are not very different.  Equation (\ref{fp1}) is therefore often 
replaced by a simple diagonal approximation
\begin{equation}
\sigma_{\gamma\,p}(\nu,Q^{2}) = 
\int_{m_{0}^{2}}^{\infty} dm^{2} \frac{\rho(\nu,m^{2})}{
(m^{2}+Q^{2})^{2} } \; ,
\label{fp2}
\end{equation} 
where  \( \rho(\nu,m^{2}) \) is an effective quantity, meant to represent the
more complicated structure on average\footnote{ If (\ref{fp2}) were exact, in
the context of scaling in $e^+ \, e^-$ annihilation and deep inelastic
scattering it would imply that the total cross-sections $\sigma_m$ for  
states $m$
scattering on nucleons decreased approximately as $m^{-2}$. This 
counter-intuitive result is sometimes known as the Gribov paradox. It implies
that the mean free paths in nuclear matter increase like $m^2$, so that
shadowing would die away at large $Q^2$ at fixed $x$. These problems do not 
occur in
\lq \lq off-diagonal models'' of the form (\ref{fp2}), when the effective 
quantity $\rho(\nu, m^2)$ results from destructive interference between
diagonal and off-diagonal terms in (\ref{fp1}).( Again, see \cite{Shaw93,
Shaw89, Review} and references therein.)}. 
Single hadron-like behaviour of the
intermediate states can be incorporated by assuming a Regge-type energy
dependence 
$$
 \rho^P(\nu, m^2)   = f_P(m^2) \nu^{\alpha_P - 1}
$$
for the dominant diffractive contribution, where $f_P(m^2)$ is a  
smoothly varying 
function of mass chosen to lead to approximate scaling at large $Q^2$. In 
particular, if we assume
\begin{equation}                  
 \rho^P(\nu,m^{2}) = a \nu^{\alpha_P - 1} (m^{2})^{1 - \alpha_P} 
\label{fp3}
\end{equation}
this gives
\begin{equation}
F_{2}^{P} = 
A \; x \, \nu^{\alpha_{P}} \; \; \int_{m_{0}^{2}}^{\infty} dm^{2}
\frac{(m^{2})^{1-\alpha_{P}}}{(m^{2}+Q^{2})^{2} } 
\label{fp4}          
\end{equation}                                                  
for the soft pomeron contribution to the structure function,  where 
$$ 
A = \frac{M \, a}{2 \, \pi^2 \alpha}    \, .
$$
This formula obviously embodies single hadron-like behaviour $\alpha_P \approx
1.08$ for all
intermediate states. Here we  
modify it by adjusting  the contributions from $m > m_J$ and $l > l_C$, 
which is meant to roughly characterize the
region in which single hadron-like  behaviour has given way to 
a more complicated behaviour. This behaviour is unknown, so we parameterize
it by the simplest possible generalization of
(\ref{fp3}), in which $A, \; \alpha_P$ are replaced by new parameters 
 $ \tilde{A}, \; \alpha_P^{\prime}$ to allow for a different magnitude and
energy dependence. 
In this way we arrive at a representation of the form
  \begin{eqnarray}
  \lefteqn{F_2^P =  }  \nonumber \\
  & & A x \nu^{\alpha_P}  \int_{m_0^2}^{\infty} dm^2 
\frac{(m^2)^{1- \alpha_P}}{(m^2 + Q^2)^2}
 [\theta (m_J^2 - m^2) +  \theta (m^2 - m_J^2)  \theta (l_C - l)] 
\nonumber \\
  & & \mbox{} + \tilde{A} x \nu^{\alpha_P^{\prime}}
\int_{m_0^2}^{\infty} dm^2 
\frac{(m^2)^{1- \alpha_P^{\prime}}}{(m^2 + Q^2)^2}
 \theta (m^2 - m_J^2)  \theta (l - l_C) \; .    \label{eq:SF1}
\end{eqnarray} 
For data fits, this is conveniently rewritten as
\begin{eqnarray}
  \lefteqn{F_2^P = \theta (m_C^2 - m_J^2) \left[ Ax \nu^{\alpha_P}
\int_{m_0^2}^{m_J^2} dm^2 
\frac{(m^2)^{1- \alpha_P}}{(m^2 + Q^2)^2}  \right.} \nonumber \\
  & &\left. \mbox{} + \tilde{A} x \nu^{\alpha_P^{\prime}}
\int_{m_J^2}^{m_C^2} dm^2 
\frac{(m^2)^{1- \alpha_P^{\prime}}}{(m^2 + Q^2)^2}
 + Ax \nu^{\alpha_P} \int_{m_C^2}^{\infty} dm^2 
\frac{(m^2)^{1- \alpha_P}}{(m^2 + Q^2)^2}
 \right] \nonumber \\
  & & \mbox{} + \theta (m_J^2 - m_C^2) \left[ Ax \nu^{\alpha_P}
\int_{m_0^2}^{\infty} dm^2 
\frac{(m^2)^{1- \alpha_P}}{(m^2 + Q^2)^2}
 \right]         \label{fp7}
\end{eqnarray}
where the integrals
can conveniently approximated  by rapidly convergent series, as shown in
the Appendix, and
the  critical coherence length is replaced by an effective critical mass,
\begin{equation}
  \label{eq:mcdef}
  m_C^2(x,Q^2) \equiv   Q^2 \left( \frac{1}{xM_{p}l_{C}} - 1 \right),
\end{equation}
which, unlike $m_J^2$ (and $l_{C}$), depends on $Q^2$ and $x$.
The real photon equivalent is
\begin{eqnarray}
  \lefteqn{\sigma_{\gamma p}^{P}(\nu , 0) = \theta (2 \nu / l_{C} - m_J^2) 
\left[B \nu^{\alpha_P - 1} \left( (m_0^2)^{ - \alpha_P} - 
(m_J^2)^{ - \alpha_P}  \right) \right. } \nonumber \\
 &   & \left. \mbox{} + C \nu^{\alpha_P^{\prime} - 1} \left( 
( m_J^2)^{ - \alpha_P^{\prime} } - (2 \nu / l_{C})^{ - \alpha_P^{\prime}} 
\right) + B \nu^{\alpha_P-1} (2 \nu / l_{C})^{ - \alpha_P} \right] \nonumber \\
  & & \mbox{} + \theta(m_J^2 - 2 \nu / l_{C}) \left[ B \nu^{\alpha_P-1} 
( m_0^2)^{ - \alpha_P} \right]
\end{eqnarray}  \label{eq:Sig}
where
\begin{equation}
  B \equiv \frac{2 \pi^{2} \alpha}{2.568 M_{p}} \frac{A}{\alpha_{P}} 
\mbox{,\ \ \ \ \ } C \equiv  \frac{2 \pi^{2} \alpha}{2.568 M_{p}} 
\frac{\tilde{A}}{\alpha_P^{\prime}}.
\end{equation}
and the numerical coefficients are chosen to give the cross section 
in $mb$ for masses in $GeV$. 

In addition to the contributions associated with the hard and soft pomerons,
there is also an additional contribution 
associated with lower lying Regge poles.
This is a small correction in the region we are considering, and we  
incorporate it using the simple empirical form
\begin{equation}
F_2^R(x, Q^2) =  A_R x^{1 - \alpha_R} \left( \frac{Q^{2}}{Q^{2}+ a_R} 
\right)^{\alpha_R} \; ,
\label{fp5}
\end{equation}
proposed by Donnachie and Landshoff \cite{DL2}, with the Regge intercept
$\alpha_R = 0.55$. The total structure function 
is then given by 
\begin{equation}
F_2(x, Q^2) = F_2^P + F_2^R \; .
\label{fp8}
\end{equation}

In the \lq\lq pre-HERA"
regions  $ \nu < \nu_C \; \; ( Q^2 = 0)$ and $x > x_C(Q^2)$, (\ref{fp7})
reduces to the simpler   form (\ref{fp4}), and our parameterisation
(\ref{fp8}) can be made numerically almost identical to the empirical 
Donnachie and Landshoff parameterisation \cite{DL2}, which is known to 
give an excellent fit to photoproduction data and to structure
function data in the intermediate region $0.1 > x > 0.01$. Correspondingly, 
data in this
region effectively determines the parameters $A, m_0, A_R$ and $a_R$ in 
(\ref{fp7}, \ref{fp5}). The additional parameters 
 $ \tilde{A}, \alpha_P^{\prime}, m_J, l_C$ in (\ref{fp7}) describe the behaviour 
at very small $x$, with $m_J, l_C$ 
constrained to be a few GeV and a few fermi respectively.

\section{Predictions at intermediate $Q^2$}                
                            
In 1994 \cite{MS95a}, the parameters $A, m_0, A_R$ and $a_R$ were fixed by
fitting the simple representation (\ref{fp4}, \ref{fp5}, 
\ref{fp8})) to the \lq\lq pre-HERA'' data on real photoabsorption and the 
structure function at  intermediate $x$ values $0.02 < x < 0.1$. The
additional parameters $\tilde{A}, \alpha_P^{\prime}, m_J, l_C$ in 
the final form (\ref{fp7}) were then 
determined by extending the fit to include the then recently available 
H1 data \cite{H194}  
at small $x$ and large $Q^2 > 8 \, GeV^2$; and  used to predict the 
behaviour at intermediate $Q^2$ ($ 0 < Q^2 < 8 \, GeV^2$), where no data
existed. In particular, the dramatic rise at small $x$ observed at high
 $Q^2$ was predicted to decrease rapidly and shift towards $x = 0$ as $Q^2$
becomes smaller. In this short section, we compare these predictions 
with the subsequent H1 
measurements \cite{H196}, in order to test the validity of the 
model\footnote{Since the 1994 fits were based on the 1993 H1 data at large 
$Q^2$, we initially compare their predictions at lower $Q^2$ to later
data from the same experiment. The data from ZEUS and other 
experiments will be discussed explicitly in the next section.}.

Before doing this, we consider the quality of the 1994 fits for 
$Q^2 > 8 \, GeV^2$. This is illustrated in Figure 3 for the 
Regge intercepts
\begin{equation}
\alpha_R = 0.55 \hspace{1cm} \alpha_P = 1.08  \; , 
\label{intercepts}
\end{equation}
together with both a \lq \lq conventional''  value   $\alpha_P^{\prime} = 1.27$
and a very unconventional choice $\alpha_P^{\prime} = 1.08$ for the intercept
of the long-lived, high mass states associated with the hard pomeron in this
approach.
As can be seen, this intercept  was not well determined by the existing data, 
 because of the limited $x$-range  covered. However 
the 1996 H1 data \cite{H196} are more precise and extend to smaller 
$x$-values.  Plotting the same curves against this new data in Figure 4 leads 
to a clear preference for $\alpha_P^{\prime} = 1.27$, although there
are  small discrepancies between the precise new data and the fit,  
since the parameters of the latter were determined by data of lower 
accuracy.

We now compare the 1994 predictions for the intermediate $Q^2$
region  $ 0 < Q^2 < 8 \, GeV^2$ with the 1996 H1 data, restricting
ourselves to the case 
 $\alpha_P^{\prime} = 1.27$ in the light of the previous 
discussion\footnote{For $ 3 \times  10^{-4} < x < 5 \times 10^{-3} $, 
the predictions are insensitive
to the choice of $\alpha_P^{\prime}$, but outside this range the value 1.27 
rather than
1.08 is preferred.}. Since in 1994 there was no data at all in this region 
that could be used to help determine the parameters of the fit, the curves 
shown in Figure 5 represent a genuine prediction. When taken in conjunction 
with Figure 4, they show that the $Q^2$ dependence predicted by the model
is in excellent agreement with experiment.

\section{A global analysis}

In the last section we focussed upon the 1996 H1 data at small $x$ and 
intermediate $Q^2$ because the parameters
of the 1994 predictions were fixed by  high $Q^2$ data from the 1994 H1 
experiment. Here we report the results of a simultaneous chi-squared fit to 
all  the  recent photoabsorption data in the wider  kinematic region
 $0 \leq x < 0.1$ and\footnote{We restrict ourselves to $Q^2 \le 15 \, GeV^2$
since we are interested in the transition region from real photons to
deep inelastic behaviour. Our simple model will of course break down 
eventually, since it gives exact scaling as $Q^2 \rightarrow \infty$.} 
$0 \leq Q^2 \le 15 \; GeV^2$. 
In doing so, we allow all the parameters in the diffractive term (\ref{fp7}),
 including the Regge 
intercepts $\alpha_P$ and $  \alpha_P^{\prime}$, to be determined by
 photoabsorption alone, in order to see whether the 
resulting  values accord with reasonable physical expectations. However, since
the non-diffractive term (\ref{fp5}) makes only a small contribution in 
this region, we keep the Regge intercept fixed at the value $\alpha_R = 0.55$
obtained from hadron scattering data, leaving $A_R$ and $a_R$ as variable
parameters.

The data used in the fit comprise the small $x$ data of the H1 \cite{H196},
and ZEUS \cite{ZEUS96} collaborations; the small and intermediate $x$ data
of the  E665 \cite{E665} collaboration; the intermediate-$x$
data \cite{NMC95} of the NMC collaboration; and the real photoabsorption
cross sections \cite{Caldwell73, Caldwell78, H1photo93, H1photo95, 
ZEUSphoto94}.  Obviously, we need to consider the 
consistency of the different experiments. The ZEUS and H1 data cover a similar
kinematic region, and are consistent within errors, as we shall see. In
addition, for  $x \approx 0.01$ there is data from both E665 and NMC at low
and intermediate $Q^2$, as well as high $Q^2$ data from ZEUS and H1. As can 
be seen from Fig.6, the experiments are consistent with each other, except for
the three  NMC points  at $(x, Q^2) = (0.0125, 3.26) \;  (0.0125, 4.52)$ and, 
especially, $ (0.008, 3.47)$.

The resulting fits to the various data sets are shown in Figures 7-11 and the
corresponding chi-squared contributions listed in Table 1, where
in all cases, the statistical  and point systematic errors have 
been combined in quadrature.  As can be seen, 
good fits are obtained for all data sets with the possible exception of the
NMC data, which contributes 94 towards chi-squared from 66 data points.
However it is clear from Fig. 10 that the fit is satisfactory, except for a few
points, usually  towards the edge of the kinematic range of the experiment,
which make a very large contribution to $\chi^2$. For example, the three NMC
points discussed above contribute 23 towards chi-squared; and on examining
Fig. 6, it is difficult to see how any smooth curve, which fits the rest of the
data, can do any better.

With this small caveat, we conclude that the consistency of the data and the
quality of the fit are very satisfactory. The corresponding parameter values
are given in Table 2 together with their errors\footnote{These errors are, of
course, correlated; the full error matrix is available on request.}. In
addition,  we have performed a series of seven fits in which one of the
data sets listed in Table 1 is omitted, and the remaining six are fitted,
in order check whether  the parameter values are sensitive to small changes 
in the data set. The results are summarised  in Table 3.  
As can be seen, the parameters are fairly stable against such changes, except 
for the parameters $A_R, \, a_R$ associated with the small non-diffractive
term (\ref{fp5}). These are only separately well determined\footnote{The ratio
  is well determined by real photoabsorption data.} if the NMC data, covering 
the intermediate $x$-region, are included.  

We now comment briefly on the values obtained for the parameters. In contrast 
to the 1994 fits, the intercept 
\begin{equation}
 \alpha_P^{\prime} = 1.289 \pm 0.007
\end{equation}
of the  hard pomeron is well determined, while the value 
$$
  \alpha_P = 1.059 \pm 0.007
$$ 
of the intercept of the  soft pomeron
is similar to, but perhaps slightly lower than, the value $\alpha_P = 1.08$ 
favoured by the analysis of hadronic total cross-sections \cite{DL1}
in terms of a simple  Regge pole model of the pomeron. 
The value 
$$
 m_0 = 0.68 \pm 0.01 
$$ 
is  somewhat below the $\rho$ mass, as predicted long ago by generalized 
vector dominance models with off-diagonal terms\footnote{This follows 
naturally from the destructive interference between the diagonal and
off-diagonal terms \cite{Review, FRS}.}, while  the value of the  parameter 
$$
m_J = 2.51 \pm 0.02
$$
corresponding to the transition from resonance dominance to jet-like behaviour
accords well with $e^+ \, e^-$ annihilation data. Finally the value 
for the characterictic distance over which jet-like behaviour develops for
heavy states is 
$$
 l_C  = 3.45 \pm 0.06 \; f \; .
$$
This is also very reasonable, given the usual estimate $l \approx 1$ f 
for the characteristic distance over which  light $q \bar{q}$ states
develop into vector mesons, based on, for example, the onset of 
nuclear shadowing effects in photoproduction on nuclei.

\section{Summary }

We have given a unified treatment of both real and virtual photoabsorption 
data in terms of a 
modified hadron dominance model, in which the striking enhancement observed 
at HERA at very low $x$ and high $Q^2$ is attributed to
contributions from heavy long-lived fluctuations of the incoming photon.
We have shown:

\begin{itemize}

\item  that the published predictions of the model for small $x$
and low and intermediate $Q^2 < 8 \; GeV^2$, where no data previously existed,
are  confirmed by recent data;

\item that the model gives an excellent fit to the much improved 
real photoabsorption
and proton structure function  data over the whole region 
 $0 \le x < 0.1$ and $ 0 \le Q^2 \le 15 \; GeV^2$;
 
\item that the values of the parameters of the model, determined from this real
  and virtual photoabsorption data alone, are in good agreement with physical 
expectations and other sources of information. 
 
\end{itemize}

\section{Appendix: evaluation of integrals}

Data fits were performed using the MINUIT routine to minimise chi-squared 
with respect to the parameters  of equation (\ref{fp7}). This is 
computationally quite 
involved and speed and efficiency are at a premium. For this reason the
integrals in (\ref{fp7}) were evaluated by using a convenient
series expansion rather than the normal methods of numerical quadrature.

The integrals in (\ref{fp7}) can be easily written as linear combinations of
 integrals of the form
\begin{equation}
  \int^{\infty}_{b}\frac{z^{- \epsilon}}{(Q^2 + z)^{2}}dz 
\mbox{\ \ \ \  ; \ \ } 0 < | \epsilon | < 1
\end{equation}
where the lower limits usually arise from the step functions, and can depend 
on $x$ and $Q^2$. Writing $Q^2 y = z$, 
\begin{equation}
  \int^{\infty}_{b}\frac{z^{- \epsilon}}{(Q^2 + z)^{2}}dz 
  = Q^{-2(1 + \epsilon)} J(r, \epsilon)
\end{equation}
where 
\[
J(r, \epsilon) = \int^{\infty}_{r}\frac{y^{- \epsilon}}{(1 + y)^{2}}dy
\]
and $ r = b/Q^2$.

 If $r \ge 1$, then
\begin{eqnarray}
 \lefteqn{J(r, \epsilon) = \frac{1}{r^{\epsilon}} \left\{
 \frac{1}{1 + r} + \frac{\epsilon}{r} \left\{\ln \left( \frac{1 + r}{r}
 \right) - \frac{1}{1 + \epsilon} \right\} \right.}  \nonumber \\
 & & \left. \mbox{} - \epsilon(1 + \epsilon) \sum_{n = 1}^{\infty} 
 \frac{(-1)^{n + 1}}{n(n + 1 + \epsilon) r^{(n + 1)}} \right\}. \label{eq:J}
\end{eqnarray}
If $r < 1$, then
\begin{equation}
   \int^{\infty}_{r}\frac{y^{- \epsilon}}{(1 + y)^{2}}dy = 
   \frac{ \pi \epsilon }{ \sin( \pi \epsilon)}
   - \int^{\infty}_{1/r}\frac{y^{\epsilon}}{(1 + y)^{2}}dy.
\end{equation}
Hence
\begin{equation}
  J(r, \epsilon) =  \frac{ \pi \epsilon }{\sin( \pi \epsilon)} 
- J(1/r, - \epsilon)
\end{equation}
and the series for $r > 1$ in (\ref{eq:J}) can be used. These series have  
two advantages ensuring speed and accuracy of computation: 
\begin{itemize}
\item the series are alternating so, provided  the number of terms is not too 
large in view of the numerical precision used, the error in the expansion is 
known  and the degree of accuracy easily controllable;
\item the expansion variable is the ratio of the integrand lower limit to 
$Q^{2}$ and for  many of the points this is sufficiently far from unity for 
the series to converge adequately after only a few terms.  
Even for a unit ratio,  the convergence is not prohibitively slow.
\end{itemize}

\begin{center}
\begin{large}
{\bf Acknowledgements}
\end{large}
\end{center}

We would like to thank Dr. J. Forshaw for several helpful discussions; and
PPARC for financial support.

\newpage

\begin{center}
{\bf Tables}
\end{center}

\begin{table}[htb]
\begin{center}
  \leavevmode
  \begin{tabular}{l|l|c |r}
    \multicolumn{2}{c|}{Data source} & Number of points & $\chi^2$ \\ \hline
    Structure Function data & H1 & 64 & 35.7 \\
    & ZEUS & 47 & 47.1 \\
    & NMC & 66 & 93.9 \\
    & E665 & 77 & 74.0 \\
    Real photo-absorption & Caldwell'73 & 9 & 8.8 \\
    & Caldwell'78 & 30 & 33.8 \\
     & HERA & 3 & 2.3 \\ \hline
    \multicolumn{2}{c|}{All data} & 296 & $ 295.6 $ \\ \hline
  \end{tabular}
  \caption{Breakdown of $\chi^2$ into contributions from different data sets 
to the global fit.}
  \label{tab:breakdown}
\end{center}
\end{table}

\begin{table}[htb]
\begin{center}
\begin{tabular}{l|c}
 \multicolumn{2}{c}{Diffractive(pomeron)} \\  \hline
 $m_0^2$      &0.466  $ \pm$ 0.008    \\
$\alpha_P$         &1.059  $\pm$ 0.007     \\
$A$            &0.629   $\pm$ 0.022    \\ 
$m_J^2$      & 6.28     $\pm$ 0.08     \\
$\alpha_P^{\prime}$    &1.289   $\pm$0.007    \\
$\tilde{A}$  &0.457    $\pm$ 0.019      \\
$l_{C}$      & 17.5   $\pm$ 0.3        \\ \hline
 \multicolumn{2}{c}{Non-Diffractive(Regge)}   \\ \hline
$A_{R}$      &0.163  $\pm$ 0.029    \\
$a_{R}$      &0.039 $\pm$ 0.009    \\
$\alpha_{R}$ &0.547(fixed)  \\  \hline
\end{tabular}
\caption{Parameter values and errors resulting from the global fit. (Natural 
units, based on the $GeV$, are used throughout.)}
\label{tab:1215}
\end{center}
\end{table}

\newpage

\begin{table}[htbp]
\begin{center}
  \leavevmode
  \begin{tabular}{c|cccc|ccc}
 & \multicolumn{7}{c|}{Omitted Data Set}   
\\ \hline
 & \multicolumn{4}{c|}{Structure Function}  & 
 \multicolumn{3}{c}{Photoabsorption} \\
   & H1 & ZEUS & E665 & NMC & Caldwell'73 & Caldwell'78 & HERA 
\\ \hline
$m_{0}^{2}$     & 0.455 & 0.465 & 0.463 & 0.508 & 0.458 & 0.463 & 0.469 \\
$\alpha_{P}$    & 1.054 & 1.059 & 1.055 & 1.067 & 1.049 & 1.050 & 1.064 \\
$A $            & 0.628 & 0.628 & 0.638 & 0.655 & 0.654 & 0.653 & 0.617 \\
$m_{J}^{2}$     & 5.823 & 6.497 & 6.303 & 6.277 & 5.945 & 5.911 & 6.339 \\
$\alpha_{P}'$    & 1.291 & 1.316 & 1.290 & 1.274 & 1.286 & 1.287 & 1.290 \\
$\tilde{A}$     & 0.432 & 0.410 & 0.454 & 0.490 & 0.458 & 0.455 & 0.451 \\
$l_{C}$         & 13.48 & 16.90 & 17.42 & 11.60 & 17.82 & 17.72 & 17.32 \\
$A_{R}$         & 0.170 & 0.164 & 0.155 & 0.012 & 0.137 & 0.138 & 0.176 \\
$a_{R}$         & 0.0459 & 0.0402 & 0.0392 & 0.00029 & 0.0382 & 
                  0.0334 & 0.0412 \\  \hline 
$\chi^{2}$      & 259 & 236 & 221 & 188 & 286 & 262 & 293 \\
No. of Points   & 232 & 249 & 219 & 230 & 287 & 266 & 293 \\ \hline   
  \end{tabular}
  \caption{Parameter and $\chi^2$ values for fits  with one data set in 
turn excluded.}
  \label{tab:varidata}
\end{center}
\end{table}

\newpage

\begin{center}
{\bf Figure Captions}
\end{center}

{\bf Figure 1} Hadronic behaviour of the photon for different values of
the coherence  length $l$ and invariant mass squared $m^2$ of the photon
fluctuations, where $l = 1/2Mx$ for $  m^2 = Q^2$. Here  $l_C$ and $M_J$ are 
the critical values of the coherence
length and fluctuation mass that separate the two types of hadronic
behaviour. Another critical coherence length $l_0 \approx 1 f$ separates
the hadronic and purely electomagnetic behaviour.

{\bf Figure 2} The hadron dominance model Eq.(\ref{fp1}).

{\bf Figure 3} The 1994 fits \cite{MS95a} to the 1994 H1 data \cite{H194} 
at $Q^2
 = 12 \;  GeV^2$, for the values $\alpha_P^{\prime} = 1.27$(dashed line) and 
$\alpha_P^{\prime} = 1.08$(solid  line). 
     
{\bf Figure 4} Comparison of the 1994 fits \cite{MS95a}
with the 1996 H1 data \cite{H196} at $Q^2 = 12 \; GeV^2$. The curves again 
correspond to $\alpha_P^{\prime} = 1.27$(dashed line) and 
$\alpha_P^{\prime} = 1.08$(solid  line).

{\bf Figure 5} Comparison of the 1994 fits \cite{MS95a}
with the 1996 H1 data \cite{H196} at $Q^2 = 1.5$ and $5 \; GeV^2$ for 
 $\alpha_P^{\prime} = 1.27$(dashed line).

{\bf Figure 6} Comparison of the data from various experiments at 
 $x=0.0125$(lower set) and $x= 0.008$(upper set), where the latter have been
scaled by a factor of two, for clarity. The E665 points for $x= 0.008$ have 
been   obtained by linearly
interpolating between $x = 0.007$ and 0.009, and the HERA data have also been
interpolated slightly from neighbouring points. The dashed(solid) lines show 
the result of the global fit described in the text for $x =0.0125(0.008)$
respectively.

{\bf Figure 7} Comparison of the global fit 
with the 1996 H1 data \cite{H196} for representative $Q^2$ values.

{\bf Figure 8} Comparison of the global fit 
with the 1996 ZEUS data \cite{ZEUS96}.

{\bf Figure 9} Comparison of the global fit 
with the E665 data \cite{E665} for representative $x$-values.

{\bf Figure 10} Comparison of the global fit 
with the NMC data \cite{NMC95}.

{\bf Figure 11} Comparison of the global fit 
with the real photoabsorption data \cite{Caldwell73, Caldwell78, H1photo93, 
H1photo95, ZEUSphoto94}.

        \newpage                                                       
        \begin{figure}[top]
        \begin{center}
        \mbox{
        \epsfysize=10cm
        \epsfbox{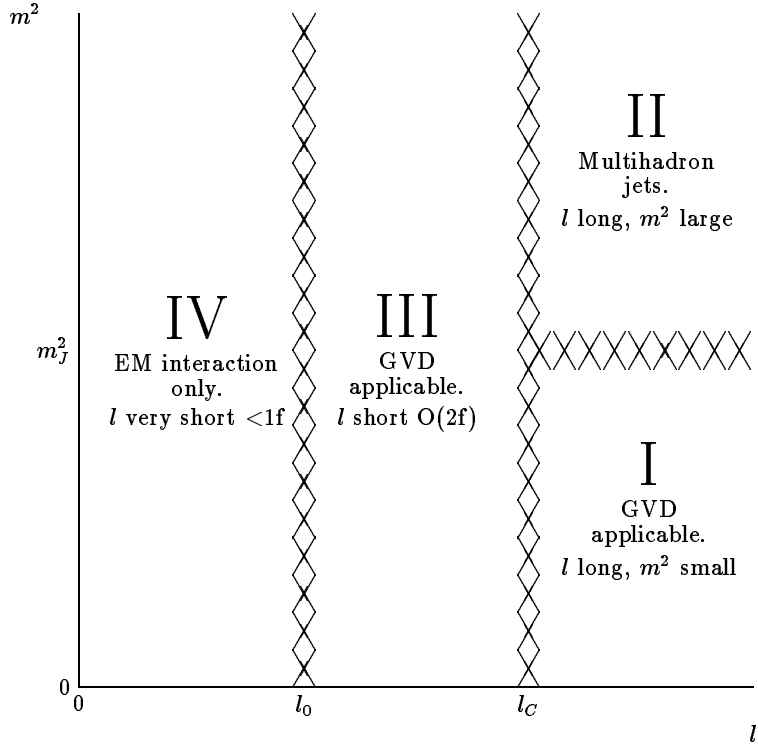}
        }
        \end{center}            
        \caption[1]{Hadronic behaviour of the photon for different values of
the coherence  length $l$ and invariant mass squared $m^2$ of the photon
fluctuations, where $l = 1/2Mx$ for $  m^2 = Q^2$. Here  $l_C$ and $M_J$ are 
the critical values of the coherence
length and fluctuation mass that separate the two types of hadronic
behaviour. Another critical coherence length $l_0 \approx 1 f$ separates
the hadronic and purely electomagnetic behaviour.}
        \end{figure}

        \newpage                                                       
        \begin{figure}[top]
        \begin{center}
        \mbox{
        \epsfysize=10cm
        \epsfbox{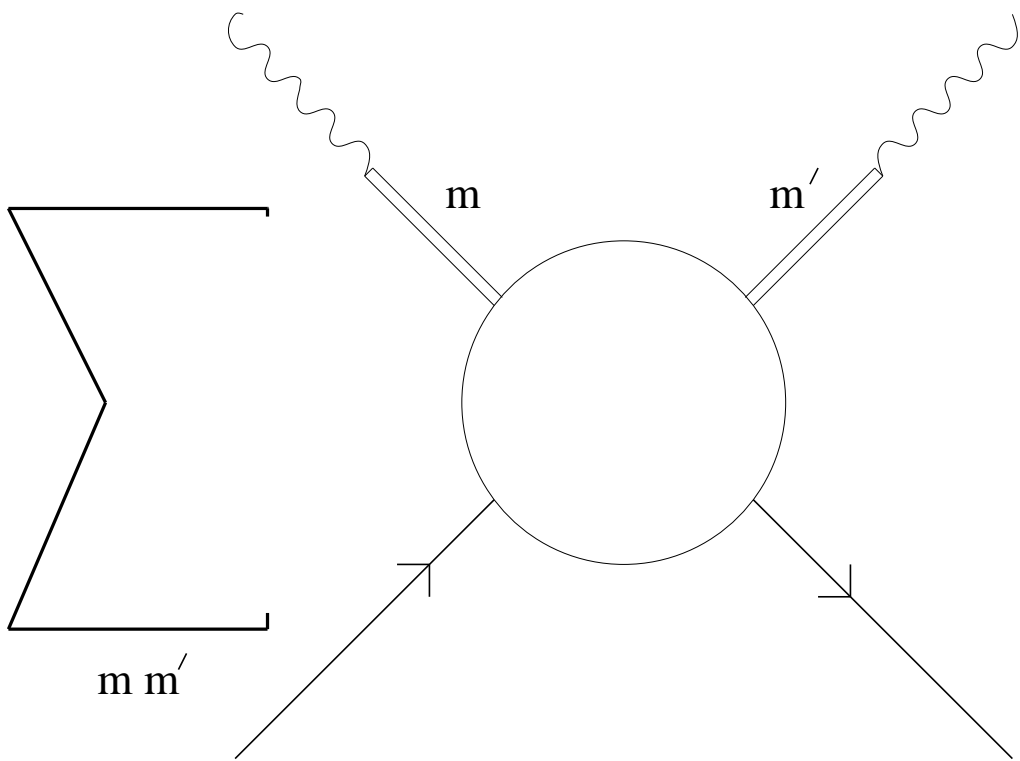}
        }
        \end{center}            
        \caption[2]{The hadron dominance model (\ref{fp1})  }
        \end{figure}         
           
        \newpage                
        \begin{figure}[t]
        \begin{center}
        \mbox{
        \epsfysize=18.0cm
        \epsfbox{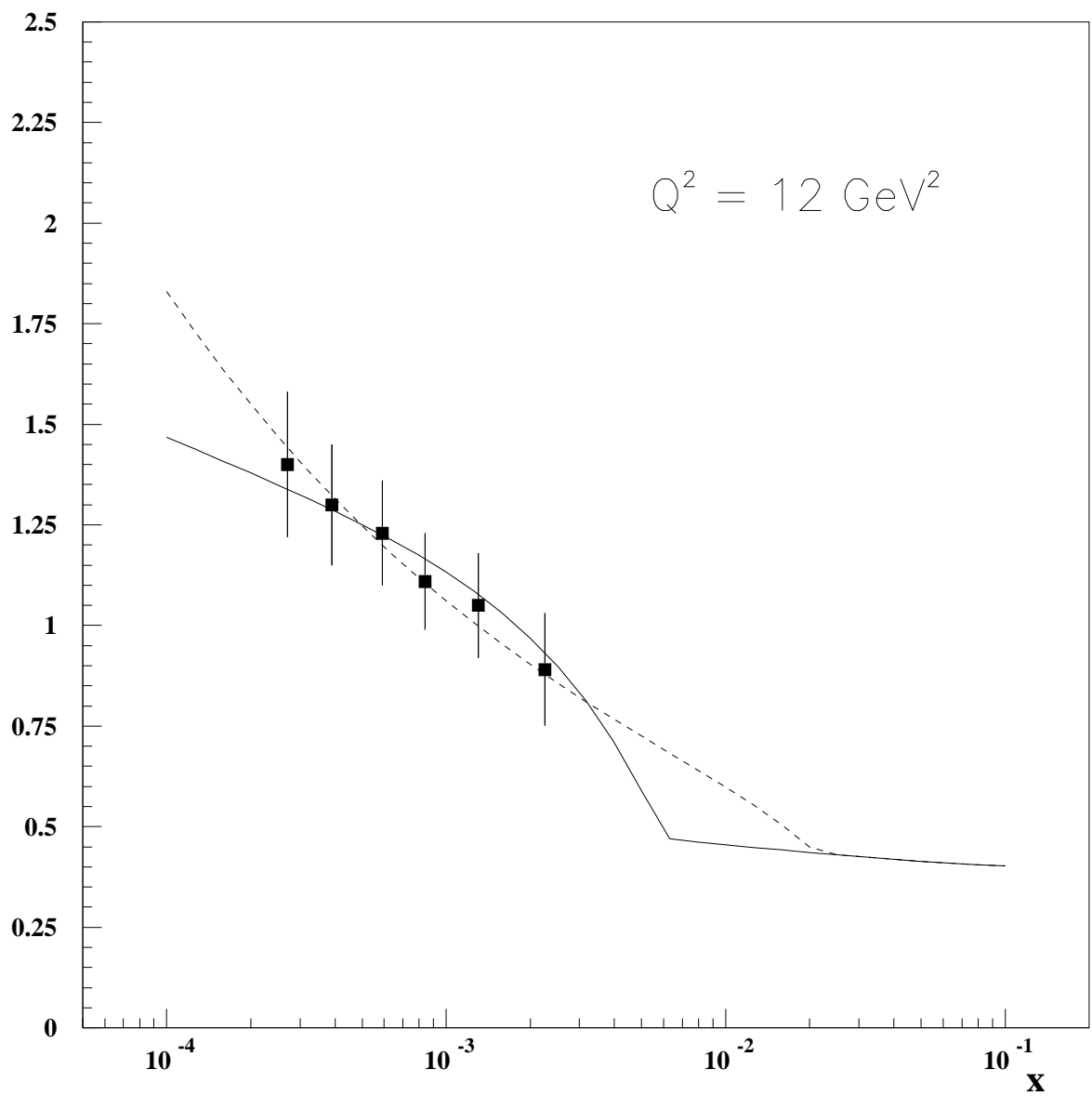}
        }
        \end{center}            
\caption[3]{The 1994 fits \cite{MS95a} to the 1994 H1 data \cite{H194} 
at $Q^2 = 12 \; GeV^2$, for the values $\alpha_P^{\prime} = 1.27$(dashed line) and 
$\alpha_P^{\prime} = 1.08$(solid  line).
                               } 
        \end{figure}

        \newpage                
        \begin{figure}[t]
        \begin{center}
        \mbox{
        \epsfysize=18.0cm
        \epsfbox{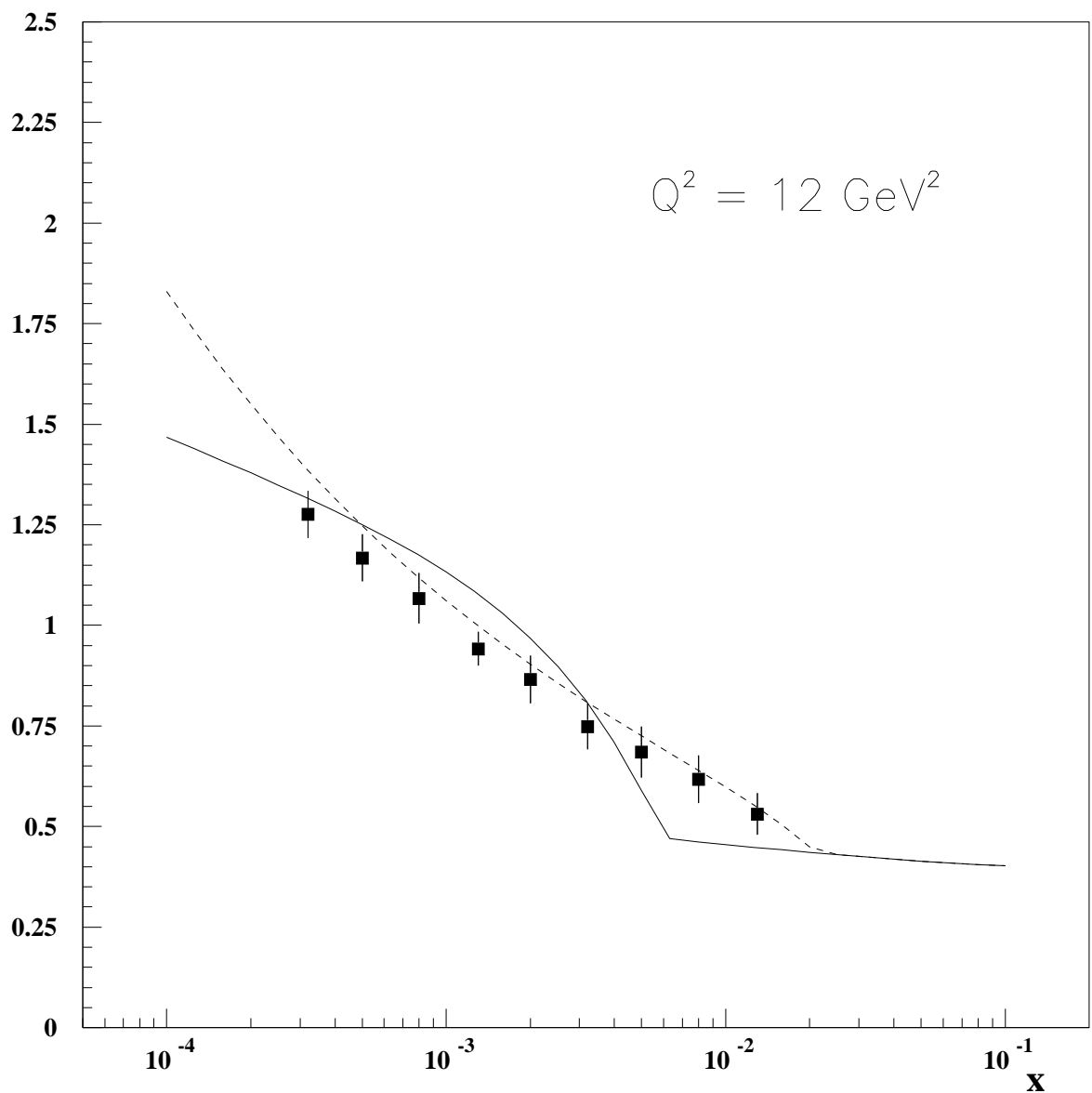}
        }
        \end{center}            
\caption[4]{Comparison of the 1994 fits \cite{MS95a}
with the 1996 H1 data \cite{H196} at $Q^2 = 12 \; GeV^2$. The curves again 
correspond to $\alpha_P^{\prime} = 1.27$(dashed line) and 
$\alpha_P^{\prime} = 1.08$(solid  line).}
        \end{figure}

        \newpage                
        \begin{figure}[t]
        \begin{center}
        \mbox{
        \epsfysize=18.0cm
        \epsfbox{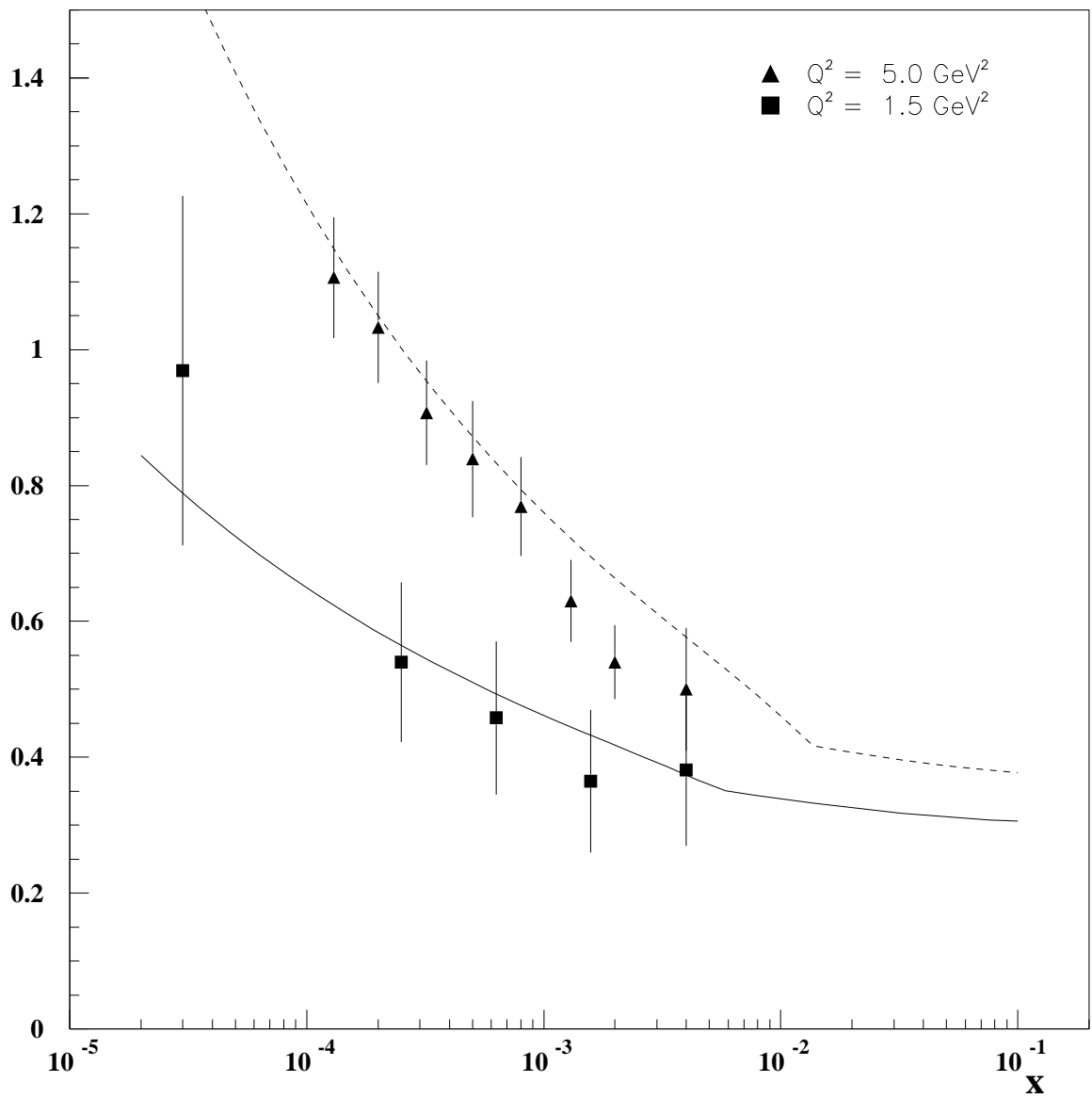}
        }
        \end{center}            
\caption[5]{Comparison of the 1994 fits \cite{MS95a}
with the 1996 H1 data \cite{H196} at $Q^2 = 1.5$ and $5 \; GeV^2$ for 
 $\alpha_P^{\prime} = 1.27$(dashed line).}
        \end{figure}

        \newpage                
        \begin{figure}[t]
        \begin{center}
        \mbox{
        \epsfysize=18.0cm
        \epsfbox{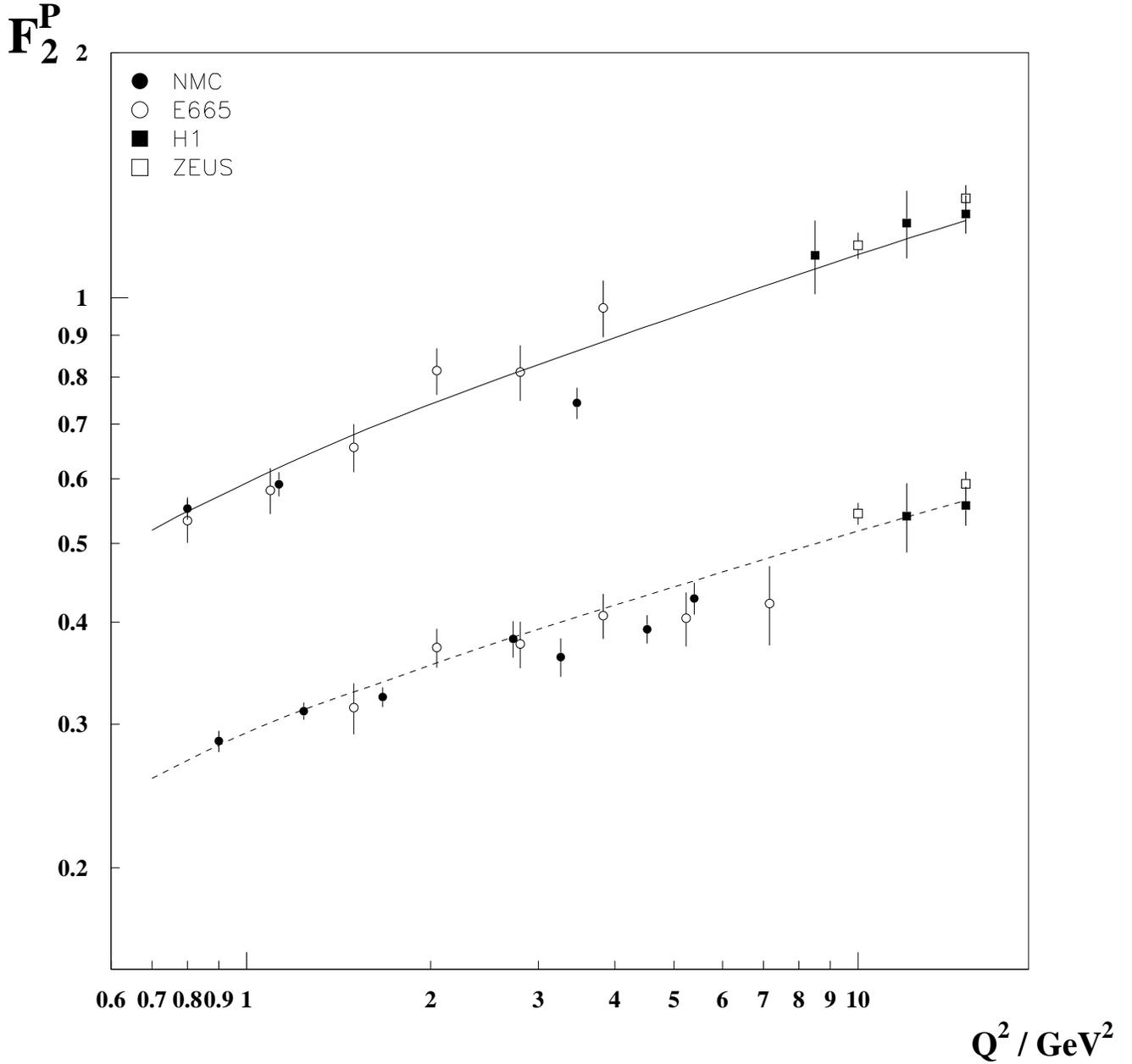}
        }
        \end{center}            
\caption[6]{Comparison of the data from various experiments at 
 $x=0.0125$(lower set) and $x= 0.008$(upper set), where the latter have been
scaled by a factor of two, for clarity. The E665 points for $x= 0.008$ have 
been   obtained by linearly
interpolating between $x = 0.007$ and 0.009, and the HERA data have also been
interpolated slightly from neighbouring points. The dashed(solid) lines show 
the result of the global fit described in the text for $x =0.0125(0.008)$
respectively }
        \end{figure}

        \newpage                
        \begin{figure}[t]
        \begin{center}
        \mbox{
        \epsfysize=18.0cm
        \epsfbox{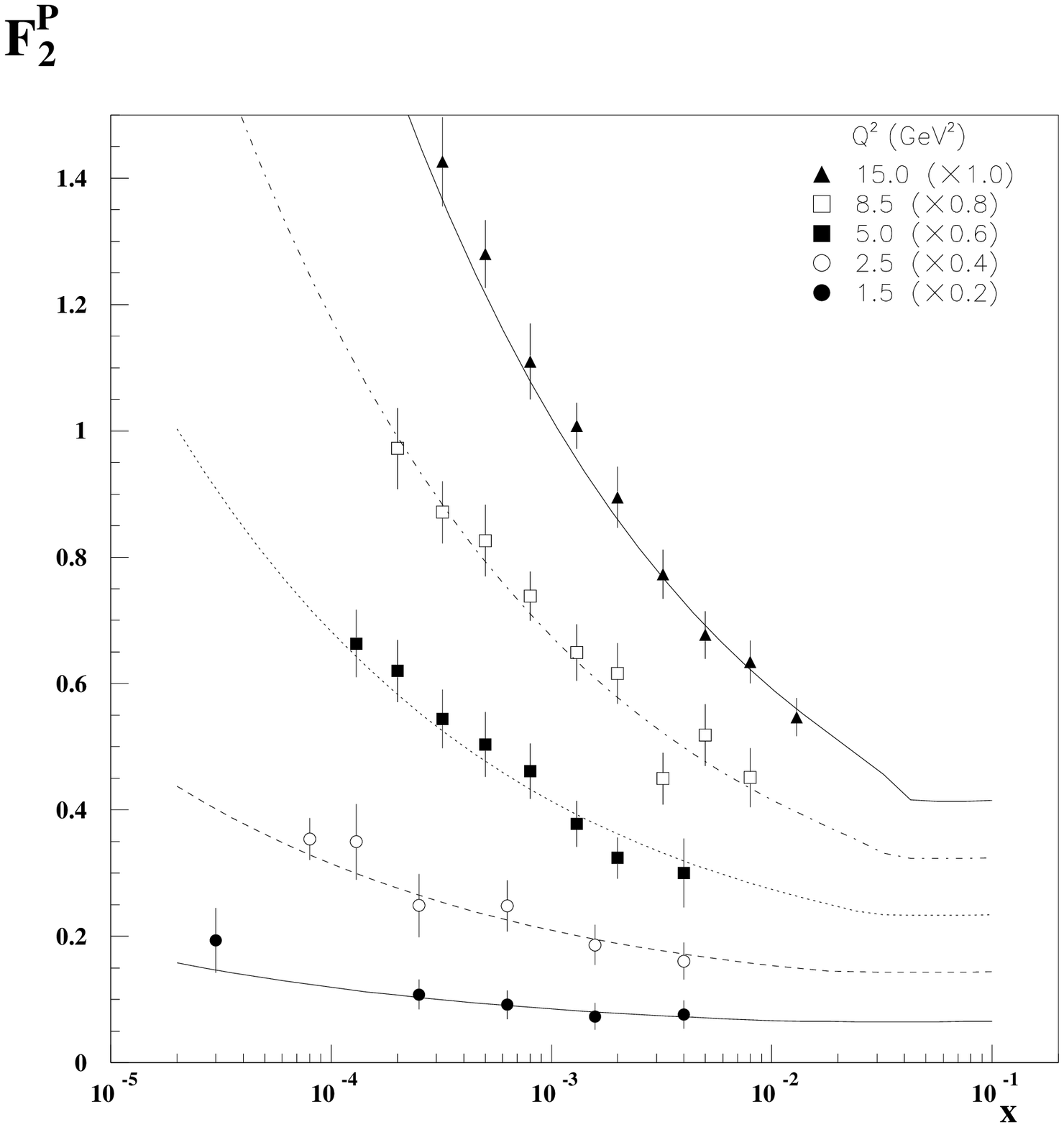}
        }
        \end{center}            
\caption[7]{Comparison of the global fit 
with the 1996 H1 data \cite{H196} for representative $Q^2$ values.}
        \end{figure}

        \newpage                
        \begin{figure}[t]
        \begin{center}
        \mbox{
        \epsfysize=18.0cm
        \epsfbox{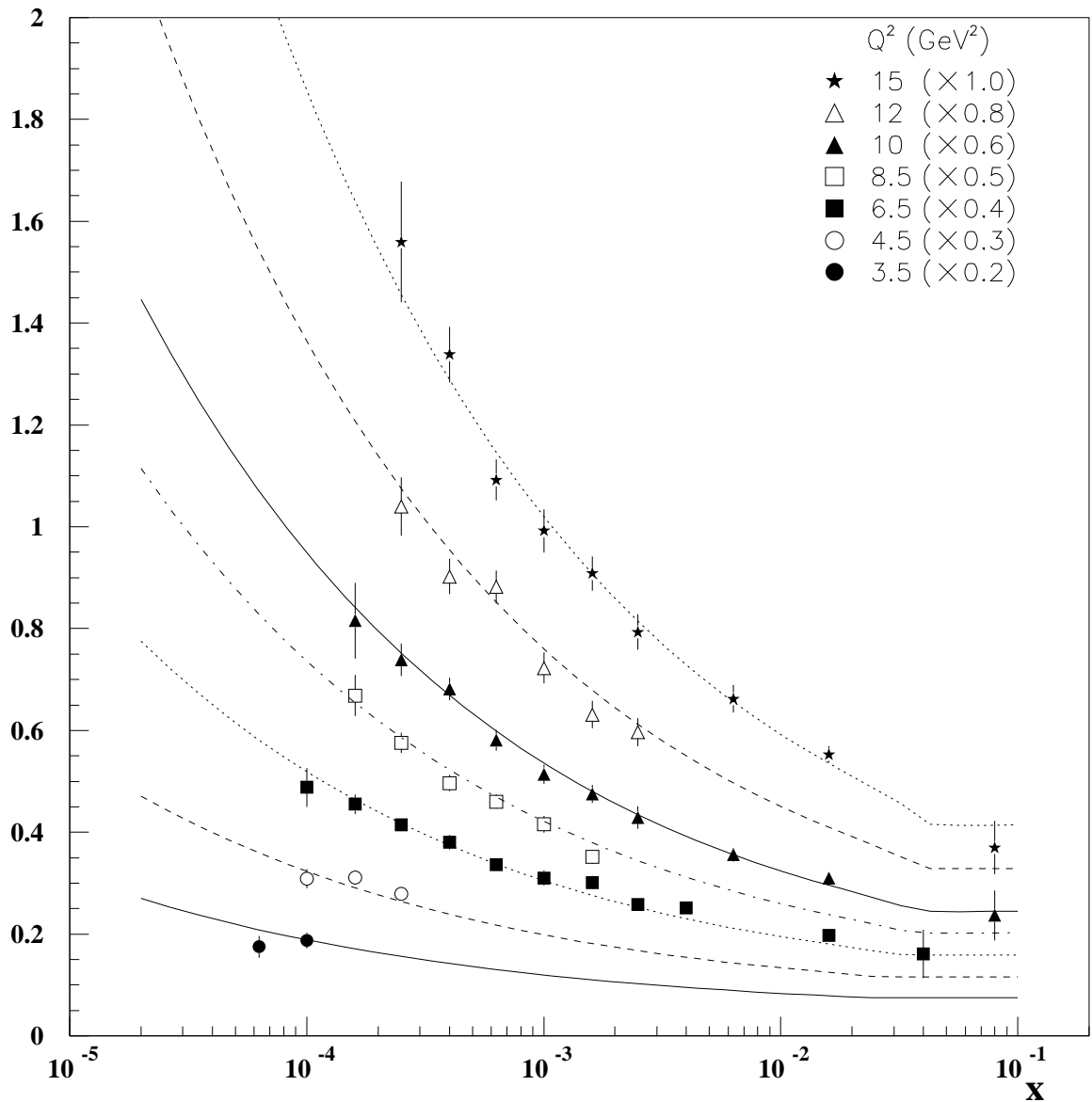}
        }
        \end{center}            
\caption[8]{Comparison of the global fit 
with the 1996 ZEUS data \cite{ZEUS96}.}
        \end{figure}

        \newpage                
        \begin{figure}[t]
        \begin{center}
        \mbox{
        \epsfysize=18.0cm
        \epsfbox{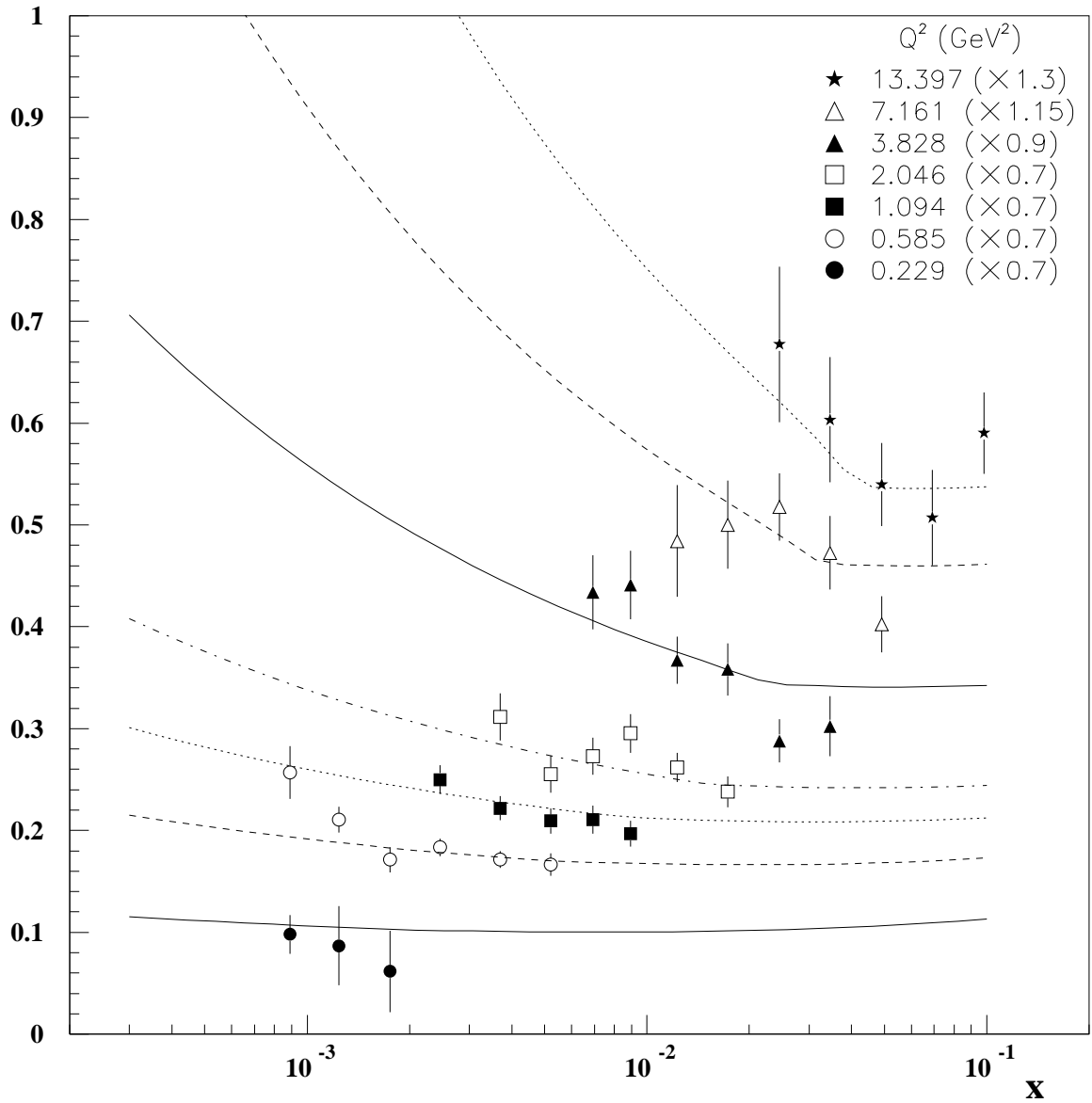}
        }
        \end{center}            
\caption[9]{Comparison of the global fit 
with the E665 data \cite{E665} for representative $x$-values.}
        \end{figure}

        \newpage                
        \begin{figure}[t]
        \begin{center}
        \mbox{
        \epsfysize=18.0cm
        \epsfbox{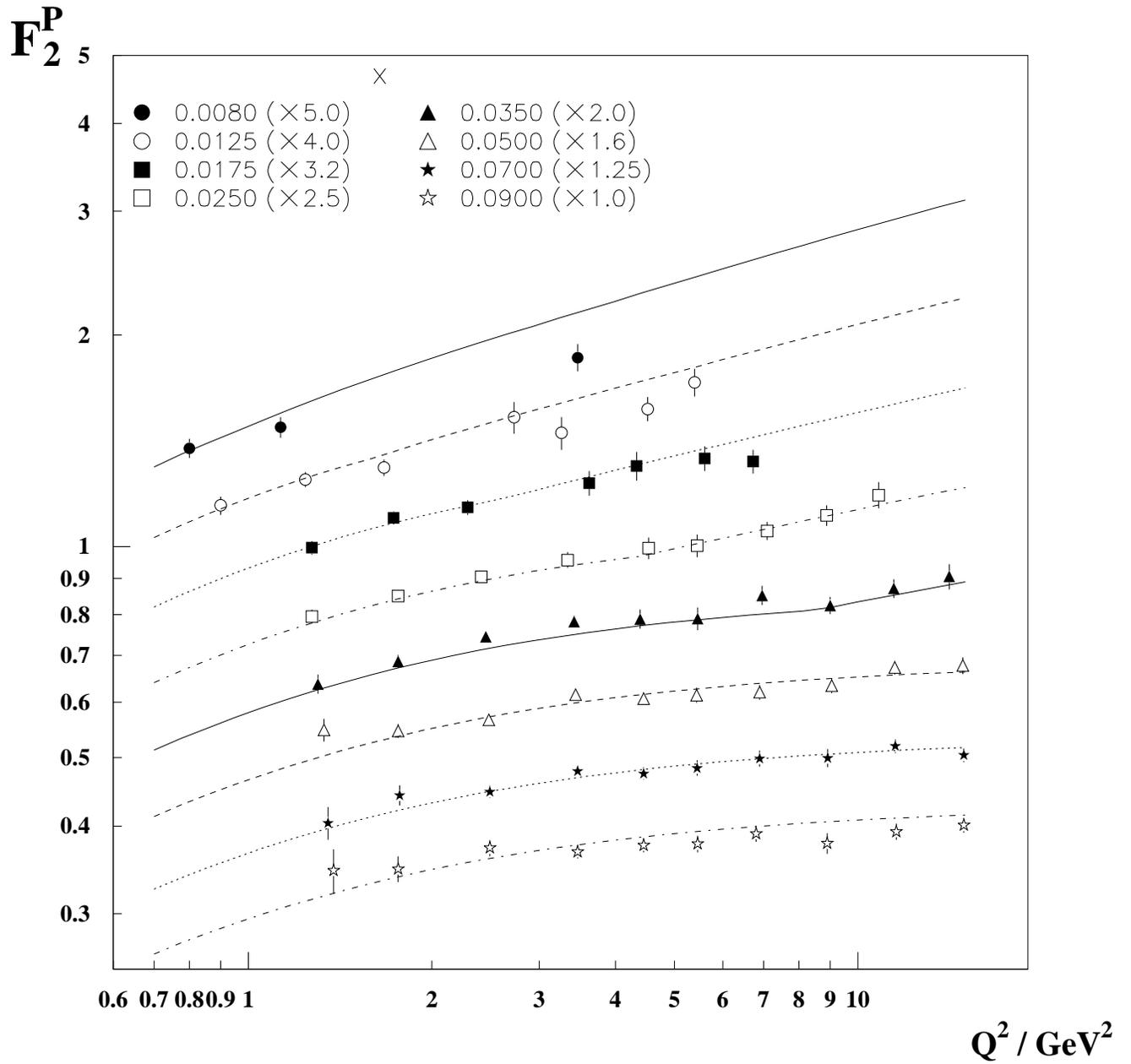}
        }
        \end{center}            
\caption[10]{Comparison of the global fit 
with the  NMC data \cite{NMC95}.}
        \end{figure}

        \newpage                
        \begin{figure}[t]
        \begin{center}
        \mbox{
        \epsfysize=18.0cm
        \epsfbox{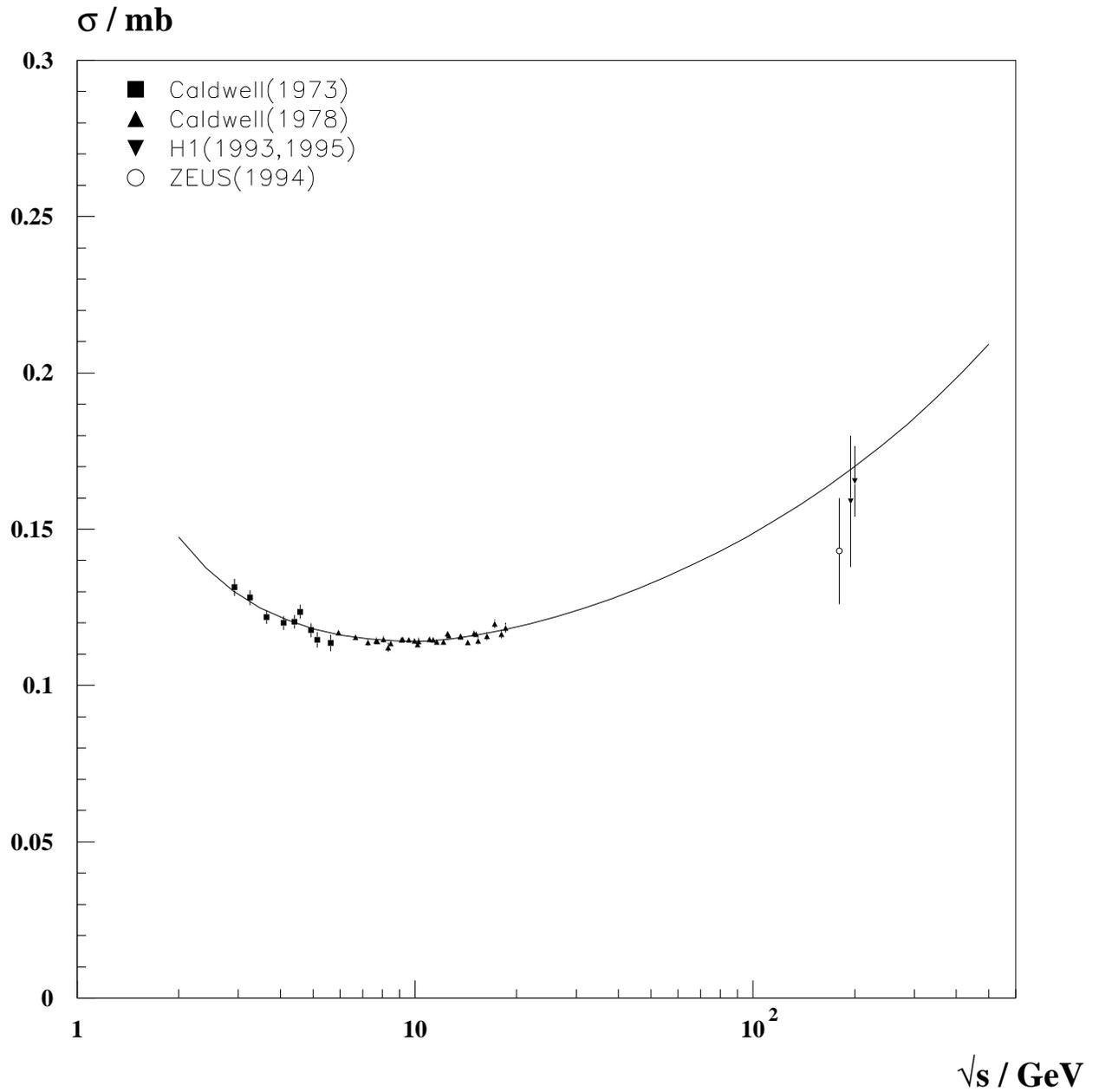}
        }
        \end{center}            
\caption[11]{Comparison of the global fit 
with the real photoabsorption data \cite{Caldwell73, Caldwell78, H1photo93, 
H1photo95, ZEUSphoto94}.}
        \end{figure}

\newpage

\end{document}